\documentclass[aps,prd,amsmath,amsfonts,a4paper,11pt,reprint,twocolumn,square,numbers,showpacs,%superscriptaddress
,floatfix,sort&compress]{revtex4-1}

\usepackage{hyperref}
\usepackage{amsmath}
\usepackage{amsfonts}
\usepackage{amssymb}
\usepackage{bm}
\usepackage{natbib}
\usepackage{graphicx}
\expandafter\let\csname equation*\endcsname\relax
\expandafter\let\csname endequation*\endcsname\relax
\usepackage{amsmath}
\usepackage[labelfont={bf,scriptsize},textfont={scriptsize},justification=RaggedRight,format=hang]{caption,subfig}
\usepackage{color}

\RequirePackage{ifpdf}
\ifpdf
\else % ordinary latex seems to include these as eps files without a problem
\fi

\def\be{\begin{equation}}
\def\ee{\end{equation}}
\def\bea{\begin{eqnarray}}
\def\eea{\end{eqnarray}}
\def\M{M_{\rm Pl}}

\def\f{\ensuremath{\mathfrak{f}}}
\def\B{\ensuremath{\mathfrak{B}}}
\def\km{\ensuremath{k_{\rm max}}}
\def\rv{\ensuremath{\rho_{\rm vac}}}
\def\rvo{\ensuremath{\rho_{0}}}
\def\ft{\ensuremath{f_{\rm tot}}}
\def\ve{\ensuremath{V_{\rm eff}}}

\begin{document}
\title{Exacerbating the cosmological constant problem \\ with interacting dark energy}

	\author{M.C.~David Marsh}
	\email{m.c.d.marsh@damtp.cam.ac.uk}
	\affiliation{Department of Applied Mathematics and Theoretical Physics, University of Cambridge,
Cambridge, CB3 0WA, United Kingdom}
		
%	\pacs{98.80.-k} 
%	\abstract{
\begin{abstract}
Future cosmological surveys will probe the expansion history of the universe and constrain phenomenological models of dark energy. Such models do not address  the fine-tuning problem of the vacuum energy, i.e.~the cosmological constant problem (c.c.p.),  but can make it
spectacularly worse.  We show that this is the case for `interacting dark energy' models in which the masses of the dark matter states depend on the dark energy sector. 
If realised in nature, these models have far-reaching implications for proposed solutions to the c.c.p.~that require the number of vacua to exceed the fine-tuning of the vacuum energy density.  We show that current estimates of the number of flux vacua in string theory,  $N_{\rm vac} \sim {\cal O}(10^{272,000})$, is far too small to realise certain simple models of interacting dark energy \emph{and} solve the cosmological constant problem anthropically. These models  admit distinctive observational signatures that can be targeted by future gamma-ray observatories, hence making it possible to observationally rule out the anthropic solution to the cosmological constant problem in theories with a finite number of vacua.  

  	\end{abstract}
	
	\maketitle
The universe expands at an accelerating rate \cite{Riess:1998cb, Perlmutter:1998np}, apparently driven by a negative pressure `dark energy' with an energy density of \cite{Ade:2015xua, Aubourg:2014yra},  
\be
\rvo \approx \left(2.3\times 10^{-3}\, {\rm eV}\right)^4 \left( \frac{\Omega_{\Lambda}}{0.69}\right) \left( \frac{h}{0.68}\right)^2 \, .
\label{eq:rhoobs} 
\ee
There are many  computable contributions to the vacuum energy density, $\rho_{\rm vac}$, but unfortunately these tend to be  comparatively large:  quantum zero-point fluctuations of a field of mass $M$  generically contributes by $\sim\frac{1}{(4\pi)^2} M^4$ after renormalisation \footnote{In effective theories with a UV cut-off $\Lambda$,   states with $M>\Lambda$ are integrated out and generically  contribute to $\rv$ by $\sim \Lambda^4$.}, which then needs to be  cancelled order-by-order in perturbation theory to an accuracy of one part in 
\be
\f_{\Lambda}=\frac{M^4}{(4\pi)^2 \rho_{\rm obs}} \, ,
\label{eq:flambda}
\ee 
to yield an effective vacuum energy compatible with  \eqref{eq:rhoobs}. 
The scale $M$ is at least as large as  
the top-quark mass $m_{\rm top} = 173$ GeV (say, if supersymmetry is realised at  the TeV scale), and may be as large as the Planck mass, $\M=2.4\times 10^{18}$ GeV, thus leading to a fine-tuning of,
\be
{\mathfrak f}_{\Lambda} =
\left\{\begin{array}{l c l}
2\times10^{53} & & M=m_{\rm top}
 \, , \\
9\times 10^{117}& & M= \M 
 \, . 
\end{array}
\right.
\label{eq:flambdaExpl}
\ee
Hence, $\rho_{\rm vac}$ is extremely sensitive to high-scale physics, and the required %delicate
 cancellations are perturbatively unstable.
This is the cosmological constant problem (c.c.p.). 

Over the coming decades, several large cosmological surveys such as DESI, LSST, Euclid and WFIRST will measure the cosmic expansion history and structure growth, and  produce stringent constraints on phenomenological models of dark energy. This will make it possible to distinguish between the simplest of these models, $\Lambda$CDM, and several more general models involving time-dependent vacuum energy densities or modifications of general relativity.  It is important to note that these models do not address nor solve the c.c.p., but in many cases make this  problem  worse by introducing additional unprotected relevant operators (e.g.~potential gradients or masses) that need to be fine-tuned, and/or by requiring that $\rho_{\rm vac} \ll (10^{-3}~{\rm eV})^4$ to make room for other mechanisms to generate an accelerated expansion.

In this paper, we illustrate this point by considering models of `interacting dark energy' in which the dark matter fields are coupled to the dark energy sector, leading to varying dark matter masses. Such models may a priori appear quite natural  as dark matter/dark energy couplings are 
not in general forbidden,  and suitable interactions may potentially explain the cosmic coincidence: why $\Omega_{\Lambda}/\Omega_m \sim {\cal O}(1)$.  Interacting dark energy has been shown to be compatible, or even favoured, by current observations \cite{Wands:2014, Abdalla:2014cla, Murgia:2016ccp,  Wang:2016lxa, Costa:2016tpb, Nunes:2016dlj}. For  reviews see \cite{Bolotin:2013jpa, Wang:2016lxa}.

We here show that  %interacting dark energy models 
cosmological models with varying dark matter masses 
are far from natural and can in fact 
%increase the fine-tuning of the 
%vacuum energy density
 exacerbate the c.c.p.~to fantastic levels. The reason is simple: the vacuum energy density depends on any 
 %dark matter
  mass, $M$, and even a small variation $\delta M$  in the spectrum can induce an enormous variation in the vacuum energy density,
 \be
 \delta \rho_{\rm vac} \sim \delta M M^3 + \ldots %{\cal O}\left(  \left( \frac{\delta M}{M}\right)^2 M^4\right) \, . 
 \ee
Ensuring  that the scalar potential of the dark energy field is flat enough to give a viable cosmology then requires 
 additional, %severely
  fine-tuned cancellations. %, beyond that usually associated with the cosmological constant problem. % in the dark energy effective potential.   
  We here show that the total fine-tuning, 
  \be
  f_{\rm tot} = f_{\Lambda} f_{\delta M} \, ,
  \ee
  is unbounded from above and may be as large as $10^{10^{10}}$ in simple examples. We note that this argument is not specific to interacting dark energy models but applies more generally to any model with cosmologically varying fundamental parameters (cf.~\cite{Banks, Donoghue:2001cs}). 

Whatever the ultimate solution of the c.c.p.~is, it is not expected to set  \emph{all} potential energies to zero, 
and hence, the additional fine-tuning from varying masses  raises an additional challenge for any proposed solution of the problem. 
%as this would raise additional challenges in particle physics and cosmology, where the Higgs potential and  inflaton potential play crucial roles. 
Here, we point out that 
the extreme unnaturalness of these models,
%
%Extremely unnatural models, i
if realised in nature, would have significant conceptual  implications, in particular for %proposed 
anthropic solutions to the c.c.p., as we now discuss.  

In theories with 
%applicable to theories 
a very large number of vacua (say, $N_{\rm vac}\gg \f_{\Lambda}$) %that scan %the effective vacuum energy density
between which 
$\rho_{\rm vac}$ takes on different vacuum expectation values, 
it is plausible that 
a small subset of the vacua  have $\rho_{\rm vac} \lesssim \rho_{\rm vac,\, obs}$ by chance. Assuming a mechanism that could realise each of these vacua with some probability,  
the c.c.p.~may be  solved through environmental selection, as solutions with $\rv \gg \rvo$ would not permit cosmological structure formation or intelligent life, and hence have a vanishing probability of being observed \cite{Weinberg:1987dv, Barrow:1988yia, Weinberg:1988cp, Martel:1997vi}.

In this context, any dark energy model 
that 
requires
additional tuning  beyond that of the 
c.c.p.~is disfavoured on statistical grounds \cite{Bousso:2007gp}. Here, 
%and %
%Theories with varying  masses or `constants'  are
%, while not known to be favoured on theoretical grounds,  
%in principle compatible with an anthropic solution to the c.c.p.~
we note that such models cannot  be expected to be realised 
at all unless $N_{\rm vac} \gg \ft$, and hence, a simultaneous realisation of certain interacting dark energy models {\it and} the anthropic solution to the c.c.p.~is only justified in theories with an exceptionally large number of vacua.  

 %The drastic enhancement of $\ft$ in such models 
%
%

String theory appears to have many four-dimensional vacuum solutions \cite{BP, GKP, KKLT, AD}. Whilst the 
complete vacuum structure of string theory remains a distant goal,  
%total number of vacua in string theory is not known, 
%has been proposed to realise this mechanism \cite{BP}. While the total number of string theory vacua is not known, proxies for 
%provides a framework inhich 
the number of vacua in particular constructions has been estimated, with the largest numbers %of string vacua 
arising from  
compactifications in which generalised electromagnetic fluxes wrap non-trivial cycles of the compactification manifold. Rough estimates of the number of  such 
`flux vacua' on particular compactification manifolds include, 
%
%and indicate that the total number of vacua may be very large. 
%The largest numbers of different string vacua studied to date arise from 
%
%of string theory  give rise to a large number of four-dimensional effective theories, and rough estimates of the number of supersymmetric flux vacua on particular compactification manifolds include,
\be
N_{\rm vac} \approx
\left\{\begin{array}{l c l}
{\cal O}(10^{506}) & & \text{IIB~on}~\mathbb{CP}_{1,1,1,6,9} ~ \text{\cite{Denef:2004dm}}
 \, , \\
{\cal O}(10^{272,000}) & & \text{F-theory~on}~{\cal M}_{\rm max} ~ \text{\cite{Taylor:2015xtz}}
 \, . 
\end{array}
\right.
   \label{eq:Nest}
\ee
%for F-theory  compactifications on the manifold , which was %by shear topological complexity 
Reference \cite{Taylor:2015xtz} argued that flux compactifications on  ${\cal M}_{\rm max}$ dominate the total number of flux vacua by roughly a factor of ${\cal O}(10^{3000})$. %Here, we will assume that $N_{\rm vac}$ is a very large but finite number, and we will investigate the viability of the anthropic landscape solution in the presence of highly fine-tuned `interacting dark energy'. 
 %As these vacuum number estimates exceed the required fine-tunings of equation \eqref{eq:fLambdaExpl}, it may well be possible that the c.c.p.~is solved anthropically in this theory. 
As the vacuum number estimates \eqref{eq:Nest} are larger than the required fine-tuning in equation \eqref{eq:flambda}, it is possible that string theory flux vacua admit an anthropic solution to the c.c.p.. We here show  that this is no longer the case if  certain models of interacting dark energy are realised in nature. 
 
 We finally point out that these extremely fine-tuned models may give rise to observational signals that could be targeted by upcoming experiments. Observational evidence for such a model could then be interpreted as evidence against the anthropic solution of the c.c.p.~in any theory with a  finite number of vacua.

\section{Varying {\large $\alpha$} and the c.c.p.}
We begin by illustrating  how  cosmologically varying `constants'  exacerbate the c.c.p.~by considering models in which the fine-structure constant evolves over cosmological scales. Such models have been considered in great detail in the literature (see \cite{Bekenstein:1982eu, Barrow:2001iw, Sandvik:2001rv, Donoghue:2001cs, Barrow:2001ks, Barrow:2002db, Barrow:2014vva, Uzan:2010pm} and references therein), largely motivated by the claimed detection %of a dipolar 
a variation in $\alpha$ inferred from  observations from distant quasars \cite{Webb:1998cq, Webb:2010hc, King:2012id, Wilczynska:2015una}. However,  while 
field-dependent coupling constants are commonplace in high-energy physics, quantum effects make
varying `constants'  unnatural: this point was made independently in  ref.~\cite{Banks}, which argued that  variations  $\frac{\delta \alpha}{\alpha} \gtrsim 10^{-37}$ would require additional fine-tuning of the vacuum energy, hence making quintessence models with varying constants unnatural, and in ref.~\cite{Donoghue:2001cs} (which, however, furthermore argued that  varying `constants' %physical parameters %nevertheless 
should be expected if the anthropic principle is realised in nature. The strong assumptions leading to this conclusion have not been realised in string theory).

We here review how theories with varying $\alpha$ exacerbate the c.c.p., and we extend the results of \cite{Banks, Donoghue:2001cs}  by computing the additional fine-tuning required to keep the vacuum energy sufficiently small in such theories. Hence, we take  $\alpha$ to depend on a scalar field $\chi(t, {\bf x})$ 
%from the laboratory value $\alpha_0$. 
subject to a quantum effective potential  with both an explicit and implicit dependence on  $\chi$, %$\rho = \rho(\chi, \alpha(\chi))$, 
\be 
\ve(\mu, \chi, \alpha(\chi)) = V_{0, r}(\chi) + \frac{M^4}{(4\pi)^2} \sum_{k=1}^{\infty} f_k(\mu)\, \frac{\alpha^{k-1}}{(4\pi)^{k-1}} \, ,
\ee
 where $V_{0, r}$ denotes the renormalised `classical' potential, $M$ is a large mass scale corresponding to charged particles running in loops, and $\mu$ the renormalisation group scale. %, and the coefficients $f_k$ in general depend on the renormalisation scale $\mu$.
Equation \eqref{eq:rhoobs} is then satisfied by  
imposing the renormalisation condition,
\be 
\ve(\bar \mu , \bar \chi, \bar \alpha) = \rvo \, ,
\label{eq:RGcond}
\ee
where $\bar \chi$ denotes the local  value of the field, and $\bar \alpha$  the low-energy, laboratory value of $\alpha$  at the renormalisation scale $\bar \mu$. The fine-tuning associated with equation \eqref{eq:RGcond} is just that of equation \eqref{eq:flambda}. 
 
 Now consider a variation in the field $\chi$ which induces a small, space-time dependent variation $\delta \alpha(\chi(t, {\bf x}))$
 of maximal magnitude  $\delta \alpha_{\rm m}$. 
 %sourced by the  the field excursion 
%of $\chi$ from $\bar \chi$ to $\bar \chi +
 We first consider the field variation $\delta \chi(t, {\bf x}) \leq  \delta \chi_{\rm m}$ to be small compared to the relevant cut-off scale, $\Lambda$, so that we can expand $\alpha$ as \footnote{The light field $\chi$ will act as a mediator of long-range `fifth' force and will be subject to observational constraints from absence of violations of the weak equivalence principle \cite{Carroll:1998zi}, which however, will not be important for our present discussion.},  
 $
\delta \alpha/\bar \alpha =c  \left( \frac{\delta \chi}{\Lambda}\right) 
$.
%The change in the effective potential is given by,
 %\be
%\delta \ve
%=\delta V_{0,\, r}\left(\chi\right)+
 %\frac{M^4}{(4\pi)^2} \sum_{k=1}^{\infty}  %\left( 
 %c_k \left( \frac{\delta \alpha}{4\pi}\right)^k
 %\right) 
%  \, ,
 % \label{eq:dV}
%\ee
% Hence, even rather small fractional variations in $\alpha$ then require that several terms in the expansion \eqref{eq:dV} are delicately cancelled in order for $\delta \ve$ not to exceed a maximum energy density $\rho_{\rm m}$ compatible with observations. 
%allowed by observations, this requires 
%We denote the magnitude of the variation of the fine-structure constant over the observationally relevant part of the universe by 
  The vacuum energy density is  given by, 
\be
V_{\rm eff}  =
\rvo+
\delta V_{0,\, r}\left(\chi\right)+
 \frac{M^4}{(4\pi)^2} \sum_{k=1}^{\infty}  %\left( 
 c_k \left( \frac{\delta \alpha_{\rm m}}{4\pi}\right)^k \left( \frac{\delta \chi }{\delta \chi_{\rm m}}\right)^k \, ,
\ee
where the coefficients $c_k$ generically are ${\cal O}(1)$.
$V_{\rm eff}$ should 
not to exceed some observationally inferred value
$\rho_{\rm m}$ 
(which will depend on the redshift range over which $\alpha$ varies) over the entire field range $[\bar \chi, \bar \chi+ \delta \chi_{\rm m}]$. We introduce the notation,
\be
\B = \frac{M^4}{(4\pi)^2 \rho_{\rm m}  }
\, ,~~\delta= \left(\frac{\delta  \alpha_{\rm m}}{4\pi} \right) \, ,
\ee
and note that 
%for  $c_k\sim {\cal O}(1)$, already very small field variations  $(\delta \chi/\delta \chi_{\rm m}) \sim 1/(\B \delta)$ generically induce too large contributions to the vacuum energy density.  Hence, 
ensuring the flatness of the effective  potential  requires 
%all orders of $(c\, \delta \chi/\Lambda)^k$ up to require 
fine-tuned cancellations 
between loop corrections and  terms in $\delta V_{0,r}$: for terms of  ${\cal O}(\delta \chi^{k})$,  the accuracy of the cancellation can be estimated to one part in $\B \delta^k$, requiring some amount of fine-tuning up to order 
 $k_{\rm max}={\rm floor}(\ln (\B)/\ln(1/\delta))$.
%\be
%\B\, \delta^k  \equiv \left(\frac{M^4}{(4\pi)^2 \rho_{\rm m}}  \right) \left( \frac{\delta \alpha}{4\pi}\right)^k > 1 \, ,
%\ee
%a given variation $\delta \alpha$ 
%require fine-tuned cancellations. 

In direct analogy with equation \eqref{eq:flambda}, 
the exacerbation of the c.c.p.~caused by  varying $\alpha$ is given by, %requiring that each term of \eqref{eq:deltarho} contributes 
%
%is then given by,
$
%\bea
%\f_{\delta \alpha} &=& \prod_{k=1}^{k_{\rm max}} \left( \frac{M^4}{(4\pi)^2 \rho_m(z=3)}  \left(\frac{ \alpha}{(4\pi)^{2}} \frac{\delta \alpha}{\alpha}\right)^k \right) \, ,
%\label{eq:feqn}
%\\
%&\equiv&
%\prod_{k=1}^{k_{\rm max}} \B\, \delta^k \, .
%\\
\f_{\delta \alpha} 
%&\equiv&
=
\prod_{k=1}^{k_{\rm max}} \B\, \delta^k  
%\eea
$,
and the 
%\eea
%where $k_{\rm max}$ is the smallest integer satisfying $\B\, \delta^{k_{\rm max}+1} <1$, and hence the final order of the expansion \eqref{eq:deltarho} that needs to be tuned small. 
 total fine-tuning 
 %required to solve the cosmological constant problem in the past \emph{and} present is then given by, 
 of the c.c.p.~in theories with varying $\alpha$ is then given by, 
\bea
\f_{\rm tot} &=& \f_{\Lambda} \f_{\delta \alpha} %= \frac{\rho_m(z=3)}{\rho_{\rm obs}} \prod_{k=0}^{k_{\rm max}} \B\, \delta^k 
=
r
\prod_{k=0}^{k_{\rm max}} \B\, \delta^k
%&=&r\, {\rm exp}\left(\left((k_{\rm max}+1)\left(\ln \B - \left( \frac{k_{\rm max}}{2} \right)  \ln \delta^{-1}\right) \right) \right) \nonumber \\
\\ &=& r\, \B^{(\km+1)\left(1 - \frac{\km\, \ln(\delta^{-1})}{2 \ln (\B)} \right)}
\approx r\, \B^{\frac{1}{2}\left(\km +1\right)}%\approx 28 \prod_{k=0}^{k_{\rm 
%max}} \B\, \delta^k 
\, , \nonumber
\label{eq:ftot1}
\eea
where $r= \rho_{\rm m}/\rvo$, and where in the last step we have approximated 
$\km \approx \frac{\ln \B}{\ln (1/\delta)}$.

%For a variation of the fine-structure constant given by \eqref{eq:da}, we have,
To illustrate the severity of the additional fine-tuning, we take $\delta \alpha_{\rm m}/\bar \alpha = 10^{-6}$, as motivated by the observations \cite{Webb:2010hc, King:2012id, Wilczynska:2015una}, and impose that the vacuum energy is subleading to the  matter energy density at $z= 3$ (around which time light from the distant quasars was emitted)  so that $\rho_{\rm m} = 4^3\, \Omega_{m,0}\, \rho_c$.  We then have,
\be
\delta = 6\times 10^{-10}\, , ~~~\B = 9\times10^{50} \times\left( \frac{M}{100\, {\rm GeV}}\right)^4 \, ,
\ee
giving,
\be
\f_{\rm tot} =
\left\{\begin{array}{l c l}
2\times10^{174} & & M=m_{\rm top} %= 173~{\rm GeV},
%~(k_{\rm max} = 5) 
\, , \\
3\times10^{795}& & M= \M 
%=2.4\cdot 10^{18}~{\rm GeV}
%\, ,~(k_{\rm max} = 11)
 \, ,
\end{array}
\right.
\label{eq:falpha}
\ee
which should be compared with the corresponding estimates for the ordinary c.c.p.~in equation \eqref{eq:flambda}. 

Hence, even minute variations of $\alpha$ substantially worsen the c.c.p., but the required fine-tunings of equation \eqref{eq:falpha} remain far smaller than the F-theory vacuum number estimate of \eqref{eq:Nest}. 

The linear dependence of $\delta \alpha$ on $\chi$ is not crucial to our argument, and the same conclusions could be reached for more general monomial form 
of the variation, $\delta \alpha \sim  \left(\bar \alpha\, \delta \chi/\Lambda\right)^q$ for $q\geq1$.
Subleading terms in the expansion will in general increase the fine-tuning somewhat, but will only become important if  the variation in $\alpha$ is caused by a large-field variation of the dark energy field, $\chi>\Lambda$. In this case, $k_{\rm max}\to \infty$ and $f_{\rm tot} \to \infty$.

\section{Interacting Dark Energy}
Are there models that could  make the fine-tuning of the c.c.p.~massively worse than what is possible in theories with varying $\alpha$? 
We will now show that the answer to this question is yes: in multi-field models of interacting dark energy,  keeping the  effective potential 
flat over multiple dimensional domain in field space (c.f.~$[\bar \chi, \bar \chi + \delta \chi_{\rm m}]^p$ for $p>1$)  requires  fantastic amounts of fine-tuning.  In \S\ref{sec:obs} we discuss the observational prospects of such models.

%
%within the maximal observationally allowed 
%
% within the relevant red-shift range, $\rho_{\rm m}$, 
%a general polynomial in the fields $\chi_i$ needs to be tuned to be very flat over a $p$-dimensional subspace in field space. 

\begin{figure}
  \includegraphics[width=0.35\textwidth, height=3 cm]{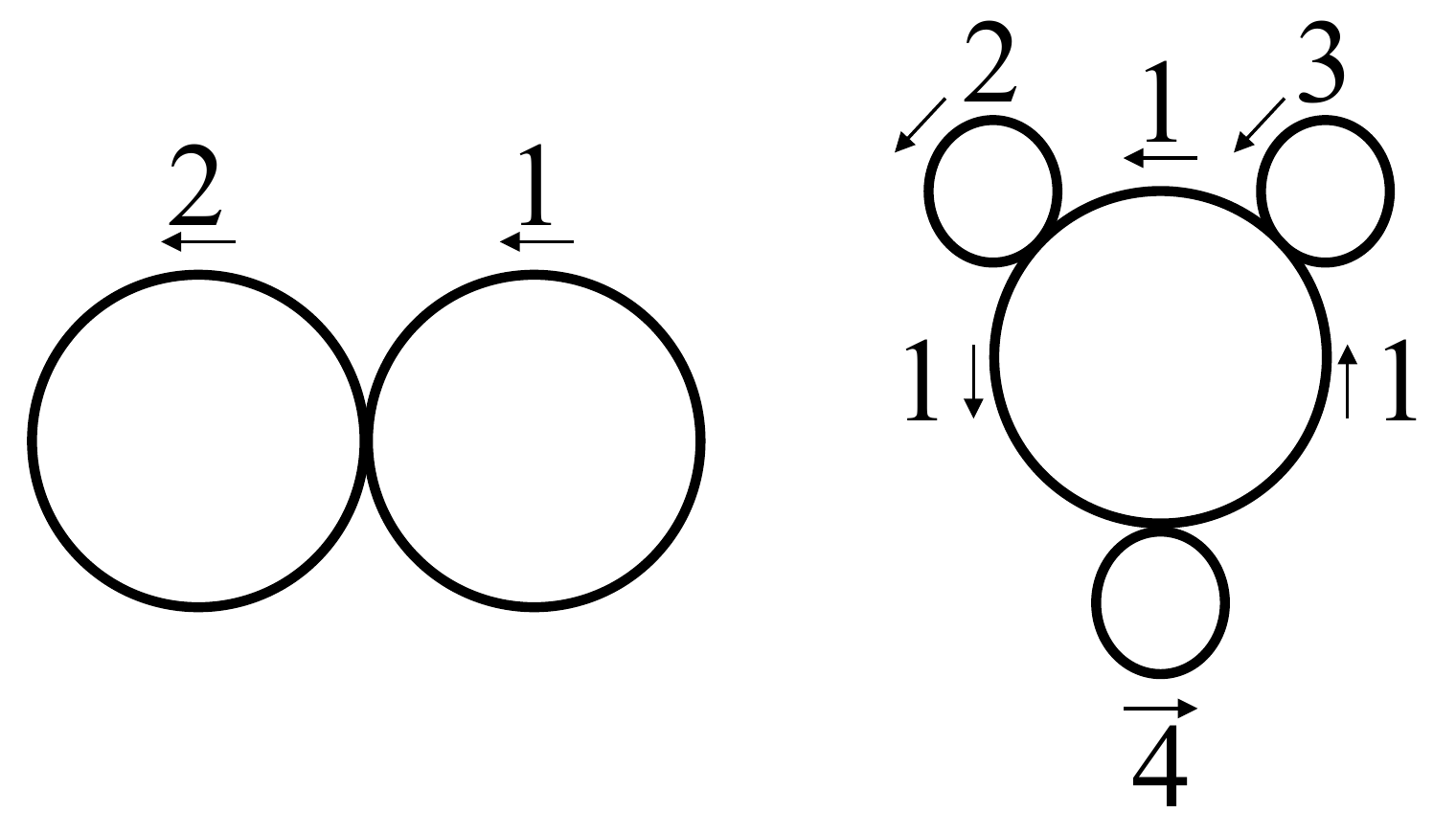}
  \caption{Contributions to the vacuum energy}
     \label{fig:Fdiag}
\end{figure}

Studies of interacting dark energy have primarily focussed on deriving CMB and large-scale structure (LSS) constraints on the form and magnitude of the energy transfer between the dark matter and dark energy sectors. Microscopic realisations of such models with any form of varying parameters exacerbate the c.c.p.~(see also \cite{D'Amico}). Here, we present a particular example of a broad class of  microscopic models that can realise a variety of phenomenological scenarios. %, while simultaneously admitting distinctive  observational signatures.
For concreteness we consider  a dark matter sector with $p$ real scalar fields, $\phi^i$, subject to the classical potential,
\bea
%{\cal L}_{\rm dm} &=& - \frac{1}{2} \partial_{\mu} \phi^i \partial^{\mu} \phi^i  - 
V_{0,\phi, r} = \frac{1}{2} m^2_i\,  (\phi^i)^2 + \frac{1}{4!} \lambda_i\,  (\phi^i)^4 + \sum_{i\neq j} \frac{1}{(2!)^2}\lambda_{ij}\, (\phi^i)^2 (\phi^j)^2 \, . \nonumber
\eea
%with renormalised masses and couplings. 
Quantum effects correct this potential at each loop order: using dimensional regularisation and $\overline{{\rm MS}}$ renormalisation,
the one-loop Coleman-Weinberg potential is given by,
\be
V_{1l} = %V_{0,r} + 
\sum_i^n \frac{m_i^4}{64\pi^2}\,  \left(\ln \left( \frac{m_i^2}{\mu^2}\right) -\frac{3}{2} \right)
\, ,
\label{eq:1l}
\ee
for 
%$L\left( \frac{m^2_i}{\mu^2}\right)=
%$ and 
where $\mu$ denotes the renormalisation scale. 
Higher loop orders induce a more complicated dependence on the masses and couplings, as we here illustrate by the processes depicted in Figure \ref{fig:Fdiag}. At two loop order, cross-terms of the form, 
\be
V_{2l} \supset \frac{\lambda_{12}}{(4\pi)^4} m^2_1 m^2_2\, L\left( \frac{m^2_1}{\mu^2}\right)\, L\left( \frac{m^2_2}{\mu^2}\right)\, ,
\ee
are induced while at four loops $V_{\rm eff}$ receives contributions of the form, 
%\bea
%\label{eq:3l}
%V_{3l} &=&\frac{1}{2} \frac{\lambda_{13} \lambda_{23}}{(4\pi)^6} m_1^2 m_2^2 \times \\
% &\times& \left(\ln\left(\frac{m_1^2}{\mu^2}\right) -1 \right)\left(\ln\left(\frac{m_2^2}{\mu^2}\right) -1\right)\ln\left( \frac{m_1^2}{\mu^2}\right)
%\, , \nonumber 
%\eea
%for the three-loop contribution of Figure \ref{fig:Fdiag}, and 
\bea
\label{eq:4l}
V_{4l} &\supset& 
 \frac{\lambda_{13} \lambda_{14} \lambda_{12}}{2\cdot 3! (4\pi)^8} \frac{ m_2^2 m_3^2 m_4^2}{m_1^2} 
 \, \prod_{i=2}^4 L\left( \frac{m^2_i}{\mu^2}\right)\, .
\eea
Here $L\left( \frac{m^2_i}{\mu^2}\right) = \ln \left( \frac{m_i^2}{\mu^2}\right) -2 $.

%for the contribution of Figure \ref{fig:Fdiag} at four loops.
%
%
%$V_{0,r}$ denotes the renormalised scalar potential
%and t

%At any loop order the vacuum energy is renormalised by %The contributions of equation \eqref{eq:CW}, \eqref{eq:3l} and \eqref{eq:4l} clearly illustrate that the vacuum energy is a complicated function of the masses of the fields. 
%To exemplify the 
%Variations in either the masses $m_i$ (with respect to some reference scale, e.g.~$\M$) or the quartic couplings, $\lambda_i,\, \lambda_{ij}$  would generate 

The dark matter masses  are now assumed to be functions of dark energy fields 
 $\chi_{\alpha} = \chi_{\alpha}(t,{\bf x})$, for   $\alpha =1 \ldots p$. %, which  evolve under the influence of . 
 The renormalisation condition of the vacuum energy  is % now 
 given by, 
 \be
 V_{\rm eff}(\bar \mu, \bar m_i) = V_{0,\phi, r}+V_{0,\chi, r} + \sum_{n=1} V_{nl} = \rvo \, ,
 \ee %at the scale $\mu_{\star}$ 
%and imposes that the effective cosmological constant is of the observationally inferred magnitude.
 with $\rvo$ as in equation \eqref{eq:rhoobs}. % and $V_{0,\chi, r}$ denoting the `classical' potential. 
 
  % and $m_i = m_i(\chi_{\alpha})$. 
 %will focus on the case in which the
 %evolution of 
For simplicity, we here take  the fields $\chi_{\alpha}$ to induce  independent variations of the masses of the form $\delta m_i/\bar m_i \sim \chi_i/\Lambda$, 
%
%\partial_{\chi_{\alpha}} (\delta m_i) \neq 0$ iff $\alpha=i$. 
 and we again assume that the variation of the field is small with respect to the cut-off $\Lambda$ (we will return to large-field models in \S\ref{sec:obs}).
 % as a models with larger field variation would be even more fine-tuned and observationally indistinguishable.
 
 %Analogously to the case of varying $\alpha$, %to keep the vacuum energy density small in this case over cosmological scales, $V_{\rm eff}$ must be tuned to be rather insensitive to the values of the masses $m_i$ and their variations. This corresponds to making 
% 
Given some fractional variations of the dark matter masses, $\delta_i \equiv \delta m_i/\bar m_i$, the change in the effective potential is   given by,
 \be
 \delta V_{\rm eff} =  \delta V_{0, \chi, r} + M^4 \sum_{ \stackrel{k_i = 0}{\text{ not all } k_i=0}}^{ \infty} \frac{c_{\underline k} }{(4\pi)^{2 l_{\underline k} }} \delta_1^{k_1} \ldots \delta_p^{k_p} \, , \label{eq:dV2}
 \ee
where $\underline k = (k_1, \ldots, k_p)$, and  $M$ denotes a suitable mass scale \footnote{For unprotected scalar masses we generically expect $M \sim \Lambda$. Here, in order not to overestimate the fine-tuning, we take $M\sim \bar m_i$. }. From equations \eqref{eq:1l}--\eqref{eq:4l}  we see that
 %in this theory,
  cross-terms between $n$ distinct masses arise at $n$-loop order, and hence, $l_{\underline k}$, which denotes the  first loop order at which the contribution with index  $\underline k$ appears, is given by $ l_{\underline k} = p - \sum_{i=1}^p \delta_{k_i}^0$ (a rather good, simple lower bound on the fine-tuning can be obtained by taking $l_{\underline k} =p$).

%can be expressed as a sum of $p$ Heaviside theta functions.

% and $l_{\underline i}$.  To %conservatively
%analytically bound the fine-tuning associated with equation \eqref{eq:dV2} from below, we may replace $l_{\underline i}$ by $i_{\rm tot} = \sum_{k=1}^p i_k > l_{\underline i}$, or alternatively  by taking $l_{\underline i} \to p$.  An upper bound %on the fine-tuning
 %is obtained by replacing $l_{\underline i} \to 1$. 

%$i_{\rm tot} =\sum_{k=1}^p i_k$. %The coefficients $c_{i_1\ldots i_n}$ so that the $p$-loop contribution that induces cross-terms between 
%We note that with this definition, if each power of $\delta$ were associated with a loop factor.  
 %
%can naturally be much larger than ${\cal O}(1)$ numbers as $V_{\rm eff}$ exhibits a complicated functional dependence on $m_i$ at already at a few loop order. 

For concreteness, we specialise to fractional variations that are independent but of the same amplitude, ${\rm max}(\delta_i) = \delta_{\rm m}$ and field variations $\chi_i \in [0, \chi_{\rm m}]$. % and define $\delta \equiv \delta_i/(4\pi)^2$ for all $i$. 
We will furthermore consider dark matter masses around $\bar m = M=100$ GeV and  take all quartic interactions $\lambda_i,~\lambda_{ij} \sim {\cal O}(1)$ so that $c_{\underline k} \sim {\cal O}(1)$. We note that this gives $\sigma/\bar m \sim 1/\bar m^3 \ll 1\, {\rm cm^2/s}$, consistent with the bound on dark matter self-interactions from the Bullet cluster \cite{Markevitch}. The  upper bound on the energy density, $\rho_{\rm m}$,  is model dependent as it is sensitive to how and when the dark energy evolves. Here, we conservatively note that supernovae data is consistent with $\Lambda$CDM \cite{Union21}, and take $\rho_{\rm m} = 10 \rvo$ for models with masses varying between $0<z \lesssim1.5$.  We then have $ \B=M^4/\rho_{\rm m} = 4\times10^{53}$.

\begin{figure}
  \includegraphics[width=0.45\textwidth, height=3.5 cm]{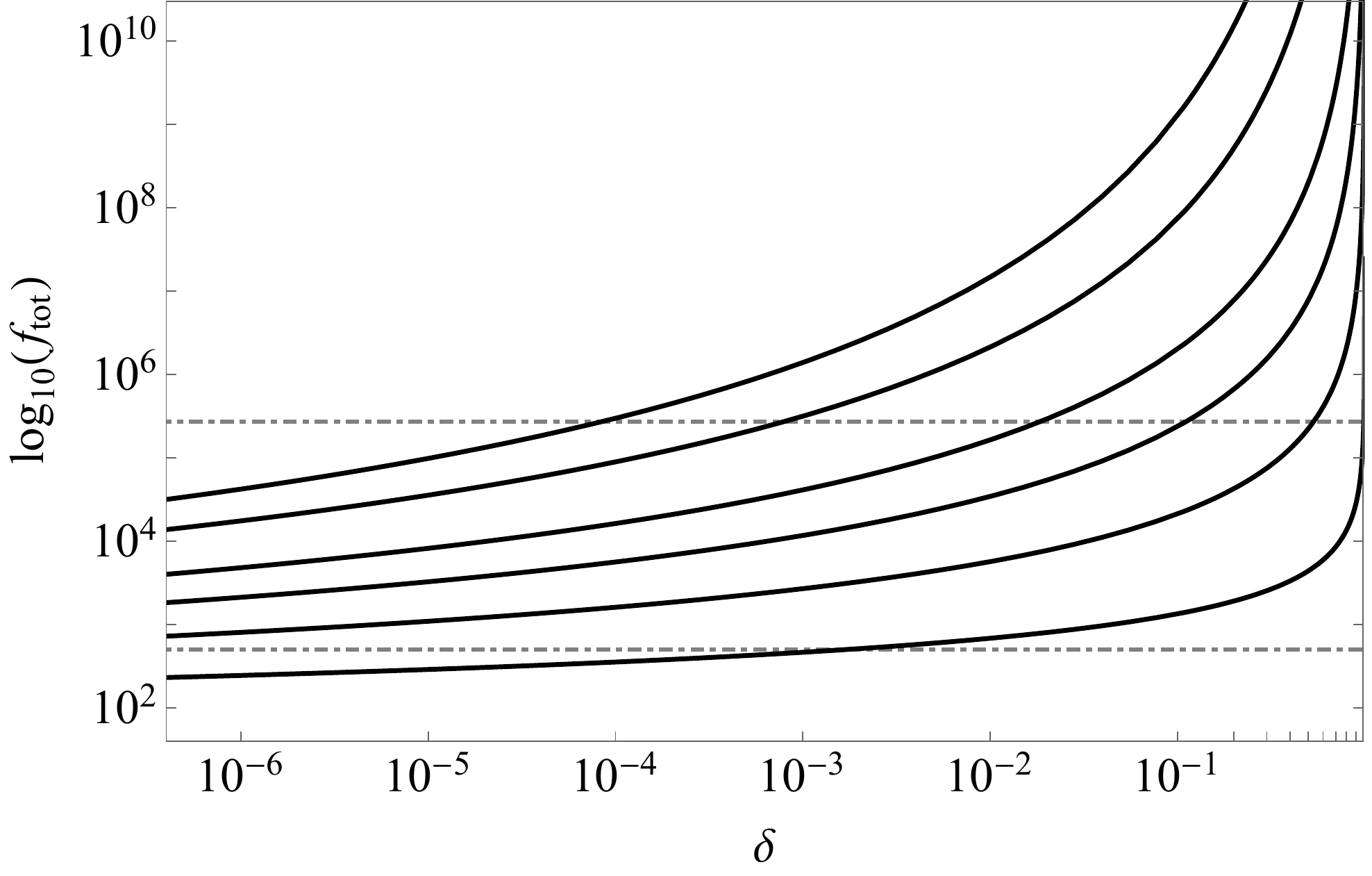}
  \caption{From bottom and up: black lines correspond to $p=1$--$4$, 6, 8; grey horizontal  lines to $\log_{10}(f_{\rm tot})=506$ and 272,000. }
     \label{fig:ft}
\end{figure}

Keeping the effective potential sufficiently flat over the entire $p$-dimensional domain $[0, \chi_{\rm m}]^p$ now requires accurate cancellations not only of terms of the form $\chi_i^k$ for some fixed species with index $i$, but of cross-terms with other fields as well. Proceeding as in the single-field case, %we find that 
the additional fine-tuning 
%due to the varying masses 
is  given by,
\bea
f_{\delta m} &\equiv& \prod_{\stackrel{k_1, \ldots, k_p =0}{{\rm not\, all\, } k_i =0}}^{\sum_i^p k_i \leq k_{\rm max}}  \B\, \frac{1}{(4\pi)^{2 l_{\underline k}}} \,\delta^{\left(\sum_{i=1}^p k_i\right)} 
%\\
%&=&
%\left(\prod_{s=1}^{{\rm min}(p,\km)}\frac{1}{(16\pi^2)^{s\, g_{p,s}}} \right)
%\left(\prod_{s =  1}^{  k_{\rm max}} 
%\left( \B\, \delta^{s}\right)^{h_{p,s}} \right)
%\, , \nonumber
\label{eq:fdmi}
\, ,
\eea
where  we take $k_{\rm max} = \ln (\B)/\ln(\delta^{-1}) + p \ln(16 \pi^2)/\ln(\delta)$.
%where now $\B= {\bar m}^4/ \rho_{\rm m}$. % while again $k_{\rm max} = {\rm floor}(\ln (\B)/\ln(1/\delta))$.
% This expression can be evaluated by noting that the degeneracy of 
%the factor $\B \delta^k$ is 
 %where the degeneracy factors are given by
 %$h_{p,k} = {p+k-1 \choose p-1}$ and
 %$g_{p,s}= { p \choose s}{k_{\rm max} \choose s-1} (k_{\rm max} +1- s)/s $.The total fine-tuning is now given by,
 % times in the product \eqref{eq:fdmi}, and hence   t
 Evaluating the product, the total fine-tuning is  given by, 
\bea 
&&f_{\rm tot} = f_{\Lambda} f_{\delta m} = 
\nonumber 
%\frac{M^4}{(4\pi)^2 \rvo}\, 
% \times \nonumber \\
%&\times& 
%\left( \frac{1}{16\pi^2}\right)^{\frac{(\km +p-1)!}{(p-1)!(\km-1)!}} 
%\B^{ \frac{(k_{\rm max} +p)!}{p! k_{\rm max} !}\left[ 1- \frac{p}{p+1} \frac{k_{\rm max} \ln(\delta^{-1})}{\ln(\B)}\right]-1} =
%\nonumber
\\
&& r \left( \frac{1}{16\pi^2}\right)^{\frac{(\km +p-1)!}{(p-1)!(\km-1)!}+1} \B^{ \frac{(k_{\rm max} +p)!}{p! k_{\rm max} !}\left[ 1- \frac{p}{p+1} \frac{k_{\rm max} \ln(\delta^{-1})}{\ln(\B)}\right]}  . ~~~~~~ %\nonumber
\label{eq:ftot}
\eea
%r \prod_{k_1, \ldots, k_p =0}^{\sum_i^p k_i = k_{\rm max}} \B\, \delta^{\left(\sum_{i=1}^p k_i\right)}   \\
%&=& r \prod_{k =0}^{ k_{\rm max}} \left(\B\, \delta^{ k}\right)^{h_{p,k}} = r\, \B^{\frac{1}{p!} \frac{(\km+p)!}{\km!} \left[ 1- \frac{p}{p+1} \frac{\km \ln (\delta^{-1})}{\ln ( \B)}\right] }\nonumber \\
%&\approx&
 % r\, \B^{\frac{1}{(p+1)!} \frac{(\km+p)!}{\km!} } \, , \nonumber
%\eea
where, again,  $r= \rho_{\rm m}/\rvo$. Equation \eqref{eq:ftot} is 
the main result of this paper and is
numerically evaluated in  Figure \ref{fig:ft}. The fine-tuning diverges as $\delta\to1$, and is extremely large already for modest  variations of multiple  masses. In particular, for $25\%$ variations of $3$ masses, 
%the c.c.p.~requires a fine-tuning of one part in 
$f_{\rm tot} \approx 10^{10^6}$,
%
%
 %$10^{10^{10}}$, far 
 exceeding the vacuum number estimates of equation \eqref{eq:Nest}.  Fine-tuning of 1 part in $10^{10^{10}}$ is required e.g.~for models with 4 masses and $\delta =0.75$. 
 
 %We conclude that there are classes of simple models of interacting dark energy that are too fine-tuned to be able to co-exist with an anthropic solution to the c.c.p.~in theories in which the number of vacua is not drastically larger than the estimates \eqref{eq:Nest}. 

\subsection{Observational prospects \label{sec:obs}}
Of crucial importance is whether the class of highly tuned interacting dark energy models discussed in this paper 
%admits  observational signatures that allows to distinguish them 
can be observationally distinguished
from less fine-tuned models. Obviously,  multiple dark matter masses can vary due to couplings to a single field $\chi$, and keeping the effective potential flat over  a 
$p=1$
 curve in field space may require much less fine-tuning than over a $p>1$ domain. Hence, simply observing multiple varying masses 
cannot be regarded as evidence for the most fine-tuned models discussed here.  Nevertheless, we will now show that  observational evidence for highly tuned models may in principle be in reach by  future experiments. 

%there exists distinctive observational signals that  models that are more fine-tuned than the vacuum number estimates \eqref{eq:Nest}

%How then, could these models be observationally established?

%The class of interacting dark energy models discussed in this paper allows for a variety of  observational signatures. E
%Most studies of interacting dark energy focus on the phenomenology of 
%energy transfer between the dark matter states and the dark energy sector, and how this affects CMB and LSS observables. 
Current and future CMB and LSS experiments will constrain the
energy transfer between the dark matter states and the dark energy sector, thereby %can be interpreted as constraints on 
 constraining
the masses, couplings, and abundances $\Omega_{\phi_i}$ in the models considered here. However, such observations are unlikely to by themselves determine 
the details of the underlying microscopic model.

The `smoking gun' signal of these models instead arise through 
very weak dark matter  couplings to photons,   leading to the decay $\phi_i \to \gamma \gamma$ that may be observable from  dark matter dense astrophysical objects. The variation in the dark matter masses would cause a variation in the energy of the photon line, $E_i(t,{\bf x}) = m_i(\chi_{\alpha}(t,{\bf x}))/2$,
 over the sky, which  could be mapped out to very high accuracy if the signal is detectable from individual galaxies or galaxy clusters. For dark matter masses of  $\sim 100$ GeV, such signals could in principle be observed by  future gamma ray observatories.
 
 To establish that the signal arises from a very fine-tuned model, we must in addition show that it corresponds to either a large-field variation ($\delta \chi/\Lambda >1$) 
 %, causing an infinite number of terms to be tuned), 
 or a  variation in a $p>1$ dimensional domain of field space. It suffices to consider a simple example to see that this may indeed %in principle 
be possible. 
 
For concreteness, we consider a  $p=3$ model in  
 which the profiles of the dark energy fields in the local universe can be approximated by $\chi_1 = \chi_0 x/L,\, , \chi_2 = \chi_0 y/L, \, , \chi_3 = \chi_0 z/L$ for some $\chi_0 \ll \Lambda$ and for some cartesian coordinates  $0< |x|, |y|, |z| < L$.  We here neglect the explicit time-dependence of the dark energy profiles as these are expected to 
 slowly on observational
time-scales, and
we again take the dark matter mass $m_i$ to depend only on $\chi_i$, so that $m_1(x),\, m_2(y)$ and $m_3(z)$. By detecting the corresponding photon lines from a large number of astrophysical objects, we can establish that a 3-dimensional domain of \emph{mass-parameter space} is sampled. Can a single field model give rise to the same observational signal? Surely so, as long as the one-parameter curve $m_i(\chi)$ covers the entire 3-dimensional mass-parameter space sufficiently densely to be consistent with  observational accuracy. However, such a single-field model is too complicated to be described by a small-field model under perturbative control, and hence, must correspond to a large field model, and hence $f_{\rm tot} \to \infty$.
 Thus, the $p=3$ small-field model with excessively large fine-tuning provides the most natural explanation of such observations.  
 
 By this example and equation \eqref{eq:ftot}, we have shown that there exists possible future observations for which models with a fine-tuning larger than the vacuum number estimates of equation \eqref{eq:Nest} provide the simplest explanation. 
 
 We note however, that the dimension of the image  of the map ${\bf x}\to \chi_i({\bf x})$  is not greater than $3$ and hence less than the dimension of the target space for $p>3$. This means that the profiles $\chi_i({\bf x})$ need to be very complicated to densely sample a $p$-dimensional domain, %, similarly,  the observational signatures of
  and we expect that such models will be  hard to 
  realise in cosmological models and, moreover, to 
  observationally distinguish from the $p=3$ case.

 %
 %acquires a non-vanishing co-dimension, hence raising an additional 
 %(though not necessarily unsurmountable)
%  challenge to observationally distinguish the case of $p=3$ to $p>3$. 
% The detection of several photon lines of dark matter origin that vary in energy %over the sky 
 %in a manner incompatible with a small  variation of a single dark energy field would indicate  multiple field dynamics (or an even more fine-tuned large-field variation), and would  provide evidence for the excessively fine-tuned models discussed here.  

\section{Conclusions}
We have shown that models in which fundamental parameters vary over cosmological scales  generically spoil the delicate cancellations of vacuum energies required by the observed smallness of the cosmological constant, in effect making the cosmological constant problem substantially worse.  Models of interacting dark energy gives a stark illustration of this point:  we have for the first time shown that there exists rather simple cosmological models with characteristic observable signatures which are too fine-tuned to be compatible with an anthropic solution of the cosmological constant problem in theories with any finite number of vacua. Hence, we have provided a concrete, positive answer to the conceptually important question of whether the string theory flux landscape, as presently understood through the estimates \eqref{eq:Nest}, is observationally falsifiable.   

We note in closing that varying `constants' can  be made natural if a mechanism dictates cancellations of the various contributions to the effective potential. Supersymmetry provides an example of such a mechanism, but supersymmetry is not realised in nature below the TeV scale and hence cannot solve the fine-tuning problem of interacting dark energy.  Other mechanisms to the same effect may in principle exist, but none have in so far been identified nor shown to be realised 
among the %low-energy
 effective  theories arising from flux compactifications for which the vacuum number estimates \eqref{eq:Nest} applies.

\section*{Acknowledgements}
I'm very grateful for stimulating conversations with  Thomas Bachlechner, John Barrow, Marcus Berg,  Raphael Bousso, Frederik Denef, Liam McAllister, and Joshua Schriffin. I acknowledge support from a Starting Grant of the European Research Council (ERC STG Grant 279617).

\bibliography{refs}

\end{document}